\documentclass[runningheads,USenglish]{llncs}
\usepackage{todonotes}
\usepackage{enumerate}
\usepackage{amsmath}
\usepackage{amssymb}
\usepackage{amsfonts}
\usepackage{wasysym}
\usepackage{latexsym}
\usepackage{xspace}
\usepackage{graphicx}
\usepackage[font=small, labelfont=bf]{caption}
\usepackage[font=small, labelfont=normal]{subcaption}
\usepackage[inline]{enumitem}

\title{Morphing Planar Graph Drawings Through 3D\thanks{This research was partially supported by MIUR Project ``AHeAD'' under PRIN 20174LF3T8.}}

\author{Kevin Buchin\inst{1}\orcidID{0000-0002-3022-7877} \and
  Will~Evans\inst{2} \and Fabrizio~Frati\inst{3}\orcidID{0000-0001-5987-8713} \and
  Irina~Kostitsyna\inst{4}\orcidID{0000-0003-0544-2257} \and Maarten~L\"offler\inst{5} \and
  Tim~Ophelders\inst{6}\orcidID{0000-0002-9570-024X} \and
  Alexander~Wolff\inst{7}\orcidID{0000-0001-5872-718X}}

\authorrunning{Buchin, Evans, Frati, Kostitsyna, L\"offler, Ophelders, Wolff}

\institute{Technische Universit\"at Dortmund, Germany,
  kevin.buchin@tu-dortmund.de
  \and
  University of British Columbia, Canada, will@cs.ubc.ca
  \and
  Roma Tre University, Italy, fabrizio.frati@uniroma3.it
  \and
  TU Eindhoven, Netherlands, i.kostitsyna@tue.nl
  \and
  Utrecht University, Netherlands, m.loffler@uu.nl
  \and
  Utrecht University \& TU Eindhoven, Netherlands, t.a.e.ophelders@uu.nl
  \and
  Universit\"at W\"urzburg, W\"urzburg, Germany}
  
\graphicspath{{figures/}}
\definecolor{defblue}{rgb}{0.121,0.47,0.705}
\definecolor{linkblue}{rgb}{0.098,0.098,0.4392}
\let\emph\relax\DeclareTextFontCommand{\emph}{\color{defblue}\em}
\usepackage[colorlinks=true, linkcolor=linkblue, anchorcolor=linkblue,
citecolor=linkblue, filecolor=linkblue, menucolor=linkblue,
urlcolor=linkblue, bookmarks=true, bookmarksopen=true,
bookmarksopenlevel=2, bookmarksnumbered=true, hyperindex=true,
plainpages=false, pdfpagelabels=true, ]{hyperref}
\usepackage[capitalise,noabbrev,nameinlink]{cleveref}

\renewcommand{\orcidID}[1]{\href{https://orcid.org/#1}{\includegraphics[scale=.03]{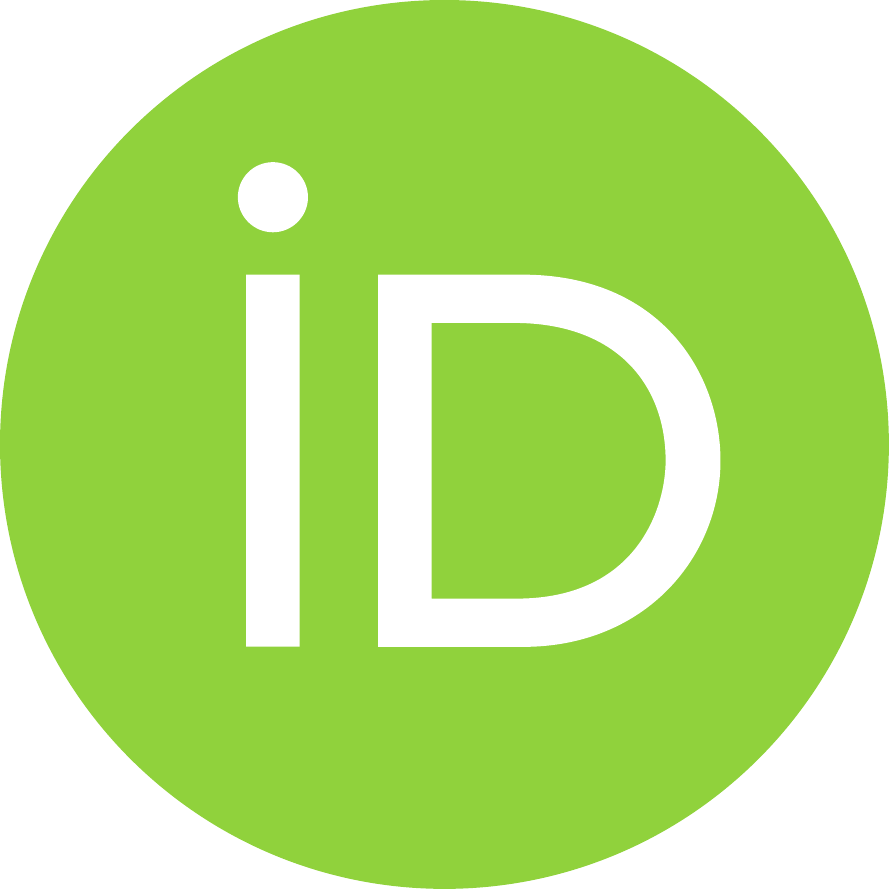}}}

\crefname{figure}{Fig.}{Figs.}
\Crefname{figure}{Figure}{Figures}
\crefname{section}{Sect.}{Sects.}
\Crefname{section}{Section}{Sections}

\newcommand{\emm}{m} 

\begin{document}

\maketitle

\begin{abstract}
In this paper, we investigate crossing-free 3D morphs between planar straight-line drawings. We show that, for any two (not necessarily topologically equivalent) planar straight-line drawings of an $n$-vertex planar graph, there exists a piecewise-linear crossing-free 3D morph with $O(n^2)$ steps that transforms one drawing into the other. We also give some evidence why it is difficult to obtain a linear lower bound (which exists in~2D) for the number of steps of a crossing-free 3D morph. 
%
\end{abstract}

\section{Introduction}
\label{sec:introduction}

A \emph{morph} is a continuous transformation between two given
drawings of the same graph.
A morph is required to preserve specific
topological and geometric properties of the input drawings. For
example, if the drawings are planar and straight-line, the morph
is required to preserve such properties throughout the
transformation. A morphing problem often assumes that the input
drawings are ``topologically equivalent'', that is, they have the same
``topological structure''. For example, if the input drawings are
planar, they are required to have the same rotation system (i.e., the
same clockwise order of the edges incident to each vertex) and the
same walk bounding the outer face; this condition is obviously
necessary (and, if the graph is connected, also sufficient~\cite{c-dplc-44,t-dpg-83}) for a morph to exist between the given drawings.
A \emph{linear} morph is a morph in which each vertex moves along
a straight-line segment, all vertices leave their initial positions simultaneously, move at uniform speed, and arrive at their final positions simultaneously. A \emph{piecewise-linear} morph consists of a sequence
of linear morphs, called \emph{steps}.
A recent line of research culminated in an algorithm by Alamdari et al.~\cite{aaccddfhlprsw-hmpgd-SICOMP17} that constructs a piecewise-linear morph with $O(n)$ steps between any two 
topologically equivalent 
planar straight-line drawings of the same $n$-vertex planar graph; this bound is worst-case optimal.


What can one gain by allowing the morph to use a third dimension? That is, suppose that the input drawings still lie on the plane $z=0$, does one get ``better'' morphs if the intermediate drawings are allowed to live in 3D? Arseneva et al.~\cite{abcddflt-pd3dm-JGAA19} proved that this is the case, as they showed that, for any two planar straight-line drawings of an $n$-vertex tree, there exists a crossing-free (i.e., no two edges cross in any intermediate drawing) piecewise-linear
3D morph between them with $O(\log n)$ steps. Later, Istomina et al.~\cite{iag-mtds3dg-WALCOM22} gave a different algorithm for the same problem. Their algorithm uses $O(\sqrt{n} \log n)$ steps, however it guarantees that any intermediate drawing of the morph lies on a 3D grid of polynomial size.


\paragraph{Our contribution.} We prove that the use of a third dimension allows us to construct a morph between any two, possibly \emph{topologically non-equivalent}, planar drawings.  Indeed, we show that $O(n^2)$ steps always suffice for constructing a crossing-free 3D morph between any two planar straight-line drawings of the same $n$-vertex planar graph; see \cref{sec:upper}. Our algorithm defines some 3D morph ``operations'' and applies a suitable sequence of these operations in order to modify the embedding of the first drawing into that of the second drawing. The topological effect of our operations on the drawing is similar to, although not the same as, that of the operations defined by Angelini et al.\ in~\cite{DBLP:journals/tcs/AngeliniCBP13}. Both the operations defined by Angelini et al.\ and ours allow to transform an embedding of a biconnected planar graph into any other. However, while our operations are 3D crossing-free morphs, we see no easy way to directly implement the operations defined by Angelini et al.\ as 3D crossing-free morphs. We stress that the input of our algorithm consists of a pair of planar drawings in the plane $z=0$; the algorithm cannot handle general 3D drawings as input. 

We then discuss the difficulty of establishing non-trivial lower bounds for the number of steps needed to construct a crossing-free 3D morph between planar straight-line drawings; see \cref{sec:lower}.  We show that, with the help of the third dimension, one can morph, in a constant number of steps, two topologically equivalent drawings of a nested-triangle graph (see \cref{fig:lower-bound}) that are known to require a
linear number of steps in any crossing-free 2D morph~\cite{aaccddfhlprsw-hmpgd-SICOMP17}.



We conclude with some open problems in \cref{sec:open}.

\section{An Upper Bound}
\label{sec:upper}

This section is devoted to a proof of the following theorem.

\begin{theorem}
  \label{thm:ub-3d}
For any two planar straight-line drawings (not necessarily with the same embedding) of an $n$-vertex planar graph, there exists a crossing-free piecewise-linear 3D morph between them with $O(n^2)$ steps.
\end{theorem}

We first assume that the given planar graph $G$ is biconnected and describe four operations (\cref{sse:operations}) that allow
us to morph a given 2D planar straight-line drawing of $G$ into another one, while achieving some desired change in the embedding. We then show (\cref{sse:biconnected}) how these operations can be used to construct a 3D crossing-free morph between any two planar straight-line drawings of $G$. Finally, we remove our biconnectivity
assumption (\cref{sse:general}).

We give some definitions. Throughout this paragraph, every considered graph is assumed to be connected. Two planar drawings of a graph are (\emph{topologically}) \emph{equivalent} if they have the same rotation system and the same clockwise order of the vertices along the boundary of the outer face. An \emph{embedding} is an equivalence class of planar drawings of a graph. A \emph{plane graph} is a graph with an embedding; when we talk about a planar drawing of a plane graph, we always assume that the embedding of the drawing is that of the plane graph.
The \emph{flip} of an embedding~$\mathcal E$ produces an embedding in which the clockwise order of the edges incident to each vertex and the clockwise order of the vertices along the boundary of the outer face are the opposite of the ones in $\mathcal E$.

A pair of vertices of a biconnected graph $G$ is a \emph{separation pair} if its removal disconnects $G$. A \emph{split pair} of $G$ is a separation pair or a pair of adjacent vertices. A \emph{split component} of $G$ with respect to a split pair $\{u,v\}$ is the edge $(u,v)$ or a maximal subgraph $G_{uv}$ of $G$ such that $\{u,v\}$ is not a split pair of $G_{uv}$. A plane graph is \emph{internally-triconnected} if every split pair consists of two vertices both incident to the outer face. 

\subsection{3D Morph Operations} \label{sse:operations}

\begin{figure}[tb]
  \begin{subfigure}[b]{.232\linewidth}
    \centering
    \includegraphics[page=1]{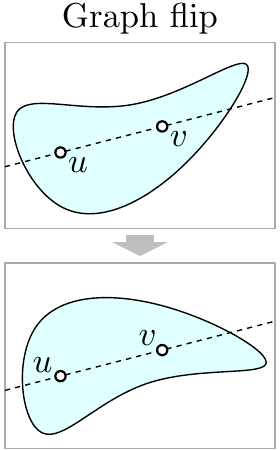}
    \caption{O1 w.r.t.\ $\{u,v\}$}
  \end{subfigure}
  \hfill
  \begin{subfigure}[b]{.232\linewidth}
    \centering
    \includegraphics[page=2]{operations}
    \caption{O2  w.r.t.\ $f$}
  \end{subfigure}
  \hfill
  \begin{subfigure}[b]{.232\linewidth}
    \centering
    \includegraphics[page=3]{operations}
    \caption{O3 w.r.t.\ $G_{2,3,4}$}
  \end{subfigure}
  \hfill
  \begin{subfigure}[b]{.232\linewidth}
    \centering
    \includegraphics[page=4]{operations}
    \caption{O4 w.r.t.\ $G_2$}
  \end{subfigure}

  \caption{The four operations that are the building blocks for our
    piecewise-linear morphs.}
  \label{fig:operations}
\end{figure}

We begin by describing four operations that morph a given planar
straight-line drawing into another with a different embedding;
see \cref{fig:operations}.

\smallskip

\noindent
{\bf Operation 1: Graph flip.}
Let $G$ be a biconnected plane graph, let $u$ and $v$ be two vertices
of~$G$, and let~$\Gamma$ be a planar straight-line drawing of~$G$.
  
\begin{lemma}
  \label{le:op1}
  There exists a $2$-step 3D crossing-free morph from $\Gamma$ to a planar
  straight-line drawing~$\Gamma''$ of $G$ whose embedding is the flip
  of the embedding that $G$ has in~$\Gamma$; moreover, $u$ and $v$ do
  not move during the morph.
\end{lemma}

We implement Operation~1, which proves \cref{le:op1}, as follows.  Let $\Pi$ be the plane $z=0$, which contains~$\Gamma$.  Let $\Pi'$ be the plane that is orthogonal
to~$\Pi$ and contains the line $\ell_{uv}$  through~$u$
and~$v$.  Let~$\Gamma'$ be the image of~$\Gamma$ under a clockwise
rotation around~$\ell_{uv}$ by~$90^\circ$.  Note that $\Gamma'$ is
contained in~$\Pi'$.  Now let~$\Gamma''$ be the image of~$\Gamma'$
under another clockwise rotation around~$\ell_{uv}$ by~$90^\circ$.
Note that $\Gamma''$ is a flipped copy of~$\Gamma$ and is contained
in~$\Pi$. Consider the linear morphs $\langle \Gamma,\Gamma'\rangle$ and $\langle \Gamma',\Gamma''\rangle$.
In each of them, every vertex travels on a
line that makes a $45^\circ$-angle with both~$\Pi$ and~$\Pi'$, and all
these lines are parallel.  Due to the linearity of the morph
and the
fact that both pre-image and image are planar, all vertices stay coplanar
during both linear morphs (although, unlike in a true rotation, the intermediate drawing size changes continuously).  In particular, every
intermediate drawing is crossing-free, and $u$ and $v$ (as well as all the points on $\ell_{uv}$) are fixed points.


\smallskip

\noindent
{\bf Operation 2: Outer face change.}  Let $G$ be a biconnected plane
graph, let~$\Gamma$ be a planar straight-line drawing of~$G$, and
let~$f$ be a face of~$\Gamma$.

\begin{lemma}\label{le:op2}
There exists a $4$-step 3D crossing-free morph from $\Gamma$ to a planar straight-line
drawing~$\Gamma'''$ of $G$ whose embedding is the same as the one of~$\Gamma$, except that the outer face of~$\Gamma'''$ is~$f$.
\end{lemma}

We implement Operation~2, which proves \cref{le:op2}, using the stereographic projection.
Let~$\Pi$ be the plane $z=0$, which contains~$\Gamma$.  Let~$S$ be a sphere
that contains~$\Gamma$ in its interior and is centered on a point in
the interior of~$f$.  Let~$\Gamma'$ be the 3D straight-line drawing
obtained   by projecting the vertices of~$G$ from their
positions in~$\Gamma$ vertically to the Northern hemisphere
of~$S$.  Let $\Gamma''$ be determined by projecting
the vertices of~$\Gamma'$ centrally from the North Pole of~$S$
to~$\Pi$.
Both projections define linear morphs:
$\langle \Gamma, \Gamma' \rangle$ and $\langle \Gamma', \Gamma'' \rangle$.
Indeed, any intermediate drawing is crossing-free since
the rays along which we project are parallel in $\langle \Gamma,\Gamma'\rangle$ and diverge in $\langle \Gamma',\Gamma''\rangle$, and there is a
one-to-one correspondence between the points in the pre-image and in
the image.  Since the morph also inverts the rotation system of~$\Gamma''$ with respect to~$\Gamma$, we apply
Operation~1 to~$\Gamma''$, which, within two morphing steps, flips~$\Gamma''$ and yields our final drawing~$\Gamma'''$.




\smallskip

\begin{figure}[tb]
  \begin{subfigure}[b]{.3\linewidth}
    \centering
    \includegraphics[page=1,scale=0.65]{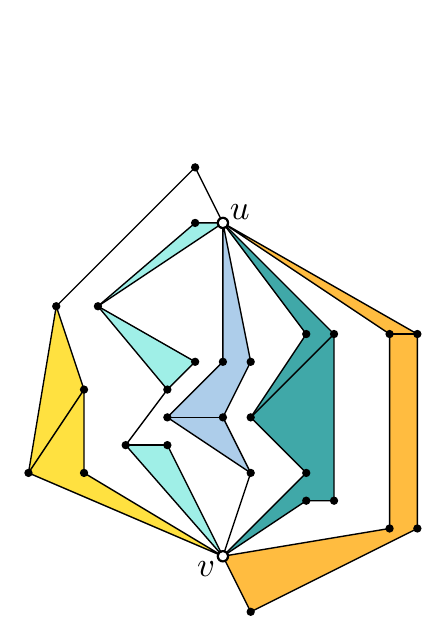}
    \caption{Drawing $\Gamma$ of~$G$.}
    \label{fig:operation-3-1a}
  \end{subfigure}
  \hfill
  \begin{subfigure}[b]{.65\linewidth}
    \centering
    \includegraphics[page=2,scale=0.65]{Operation3-1}
    \caption{Drawing $\Psi$ of $H$;
        polygon~$P_{\textrm{in}}$ is blue, $P_{\textrm{out}}$~is
        red.}
    \label{fig:operation-3-1b}
  \end{subfigure}
	
  \caption{Illustration for Operation 3 with $i=2$ and $j=4$: Construction of $\Psi$ from $\Gamma$.}
  \label{fig:operation-3-1}
\end{figure}

\noindent
{\bf Operation 3: Component flip.} Let $G$ be a biconnected plane
graph, and let $\{u,v\}$ be a split pair of~$G$.  Let $G_1,\dots,G_k$
be the split components of $G$ with respect to $\{u,v\}$. Let $\Gamma$ be a planar straight-line drawing  of $G$ in which $u$ and $v$ are incident to the outer face, as in
\cref{fig:operation-3-1a}.
Relabel $G_1,\dots,G_k$ so
that they appear in clockwise order $G_1,\dots,G_k$ around $u$, where
$G_1$ and $G_k$ are incident to the outer face of~$\Gamma$.  Let $i$
and $j$ be two (not necessarily distinct) indices with
$1\leq i\leq j\leq k$ and with the following property\footnote[1]{This is a point where our operations differ from the ones of Angelini et al.~\cite{DBLP:journals/tcs/AngeliniCBP13}. Indeed, their flip operation applies to any sequence of components of $G$, while ours does not.}: If~$G$ contains
the edge $(u,v)$, then this edge is one of the components $G_i,\dots,G_j$. Operation~3 allows us to flip the embedding of the components $G_i,\dots,G_j$ (and to incidentally reverse their order), while leaving the embedding of the other components of $G$ unchanged. This is formalized in the following.  
\begin{lemma}\label{le:op3}
There exists an $O(n)$-step 3D crossing-free morph from $\Gamma$ to a planar straight-line
drawing~$\Gamma'$ of $G$ in which the embedding of~$G_\ell$ is the flip of the embedding that $G_\ell$ has in~$\Gamma$, for $\ell=i,\dots,j$, while the embedding
of~$G_\ell$ is the same as in~$\Gamma$, for $\ell=1,\dots,i-1,j+1,\dots,k$.
The order of $G_1,\dots,G_k$ around~$u$ in $\Gamma'$ is
$G_1,\dots,G_{i-1},G_j,G_{j-1},\dots,G_i,G_{j+1},\dots,G_k$.
\end{lemma}

In order to implement Operation~3, which proves \cref{le:op3}, ideally we would like to apply Operation~1 to the drawing of the graph $G_i\cup G_{i+1}\cup \dots \cup G_j$ in $\Gamma$. However, this would result in a drawing which might contain crossings between edges of $G_i\cup G_{i+1}\cup \dots \cup G_j$ and edges of the rest of the graph. Thus, we first move $G_i\cup G_{i+1}\cup \dots \cup G_j$, via a 2D crossing-free morph, into a polygon that is symmetric with respect to the line through $u$ and $v$ and that does not contain any edges of the rest of the graph. Applying Operation~1 to $G_i\cup G_{i+1}\cup \dots \cup G_j$ now results in a drawing in which $G_i\cup G_{i+1}\cup \dots \cup G_j$ still lies inside the same symmetric polygon, which ensures that the edges of $G_i\cup G_{i+1}\cup \dots \cup G_j$ do not cross the edges of the rest of the graph.

We now describe the details of Operation~3; refer
to~\cref{fig:operation-3-1b}. We start by drawing a triangle $(a,b,c)$ surrounding $\Gamma$. Then we insert 
in $\Gamma$ two polygons $P_{\textrm{in}}$ and $P_{\textrm{out}}$ with $O(n)$ vertices, which intersect $\Gamma$ only at $u$ and $v$;
the vertices of $G_1,\dots,G_{i-1},G_{j+1},\dots,G_k$ (except $u$ and $v$) and $a$, $b$, and $c$ lie outside $P_{\textrm{out}}$;
the vertices of $G_i,\dots,G_j$ (except $u$ and $v$) lie inside $P_{\textrm{in}}$;
$P_{\textrm{out}}$ contains $P_{\textrm{in}}$;
and
the two paths of $P_{\textrm{in}}$ connecting $u$ and $v$ have the same number of vertices. 
%
%
We let $P_{\textrm{in}}$
and $P_{\textrm{out}}$ ``mimic'' the boundary of the drawing of
$G_i\cup G_{i+1}\cup \dots \cup G_j$ in $\Gamma$.

We triangulate the exterior of $P_{\textrm{out}}$; that is, we
triangulate each region inside $(a,b,c)$ and outside
$P_{\textrm{out}}$ bounding a face of the current drawing.  If this introduces a chord $(x,y)$ with respect
to~$P_{\textrm{out}}$, let $(x,y,w)$ and $(x,y,z)$ be the two faces
incident to $(x,y)$; we subdivide $(x,y)$ with a vertex and connect this vertex to $w$ and $z$.  We also triangulate the interior
of~$P_{\textrm{in}}$.  Let~$\Psi$ be
the resulting planar straight-line
drawing of this plane graph~$H$.
Let $C_{\textrm{out}}$ and $C_{\textrm{in}}$ be the cycles of $H$
represented by $P_{\textrm{out}}$ and $P_{\textrm{in}}$ in~$\Psi$, let
$H_{\textrm{out}}$ be the subgraph of $H$ induced by the vertices that
lie outside or on~$P_{\textrm{out}}$, and let~$H_{\textrm{in}}$ be
the subgraph of $H$ induced by the vertices that lie inside or
on~$P_{\textrm{in}}$.  Note that $H_{\textrm{out}}$ is a triconnected
plane graph, as each of its faces is delimited by a $3$-cycle, except
for one face, which is delimited by a cycle~$C_{\textrm{out}}$ without
chords.  Further, $H_{\textrm{in}}$ is an internally-triconnected plane graph, as each of its internal
faces is delimited by a $3$-cycle, while the outer face is delimited
by a cycle~$C_{\textrm{in}}$ which may have chords.



\begin{figure}[tb]
  \begin{subfigure}[b]{.23\linewidth}
    \centering
    \includegraphics[page=1,scale=0.5]{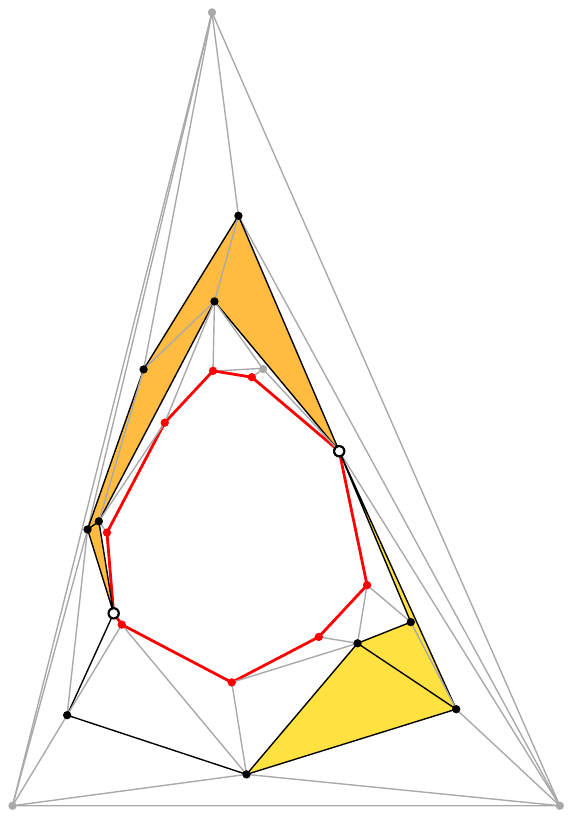}
  \end{subfigure}
  \hfill
  \begin{subfigure}[b]{.23\linewidth}
    \centering
    \includegraphics[page=2,scale=0.5]{Operation3-2}
  \end{subfigure}
  \hfill
  \begin{subfigure}[b]{.23\linewidth}
    \centering
    \includegraphics[page=3,scale=0.5]{Operation3-2}
  \end{subfigure}
  \hfill
  \begin{subfigure}[b]{.23\linewidth}
    \centering
    \includegraphics[page=4,scale=0.5]{Operation3-2}
  \end{subfigure}

  \vspace{-1ex}
    
  \begin{subfigure}[t]{.23\linewidth}
    \centering
    \caption{Drawing $\Lambda_{\textrm{out}}$ of~$H_{\textrm{out}}$.}
    \label{fig:operation-3-2a}
  \end{subfigure}
  \hfill
  \begin{subfigure}[t]{.23\linewidth}
    \centering
    \caption{Drawing
      $\Lambda_{\textrm{out}}\cup P^\Lambda_{\textrm{in}}$
      of~$H_{\textrm{out}}\cup C_{\textrm{in}}$.}
    \label{fig:operation-3-2b}
  \end{subfigure}
  \hfill
  \begin{subfigure}[t]{.23\linewidth}
    \centering
    \caption{Drawing $\Lambda$ of~$H$.}
    \label{fig:operation-3-2c}
  \end{subfigure}
  \hfill
  \begin{subfigure}[t]{.23\linewidth}
    \centering
    \caption{Drawing $\Phi$ of~$H$.}
    \label{fig:operation-3-2d}
  \end{subfigure}

  \vspace{-1ex}
              
  \caption{Illustration for Operation~3.} 
  \label{fig:operation-3-2}
\end{figure}

We now construct another planar straight-line drawing of $H$, as
follows.  Construct a strictly convex drawing $\Lambda_{\textrm{out}}$ of $H_{\textrm{out}}$, e.g., by means
of the algorithm by Hong and Nagamochi~\cite{hn-cdhpgcpg-10} or of the algorithm by Tutte~\cite{t-hdg-63}, as in~\cref{fig:operation-3-2a}. Let $P^\Lambda_{\textrm{out}}$ be
the strictly convex polygon representing $C_{\textrm{out}}$ in
$\Lambda_{\textrm{out}}$.  As in~\cref{fig:operation-3-2b}, plug a
strictly convex drawing 
$P^\Lambda_{\textrm{in}}$ of $C_{\textrm{in}}$ in the interior of
$P^\Lambda_{\textrm{out}}$ (except at~$u$ and~$v$) so that
$P^\Lambda_{\textrm{in}}$ is symmetric with respect to the line
through $u$ and $v$. This can be achieved because the two paths of~$C_{\textrm{in}}$ connecting $u$ and $v$ have the same number of
vertices and because
$P^\Lambda_{\textrm{out}}$ is strictly convex, hence the segment
$\overline{uv}$ lies in its interior, and thus also a polygon $P^\Lambda_{\textrm{in}}$
sufficiently close to $\overline{uv}$ does.  Finally, plug into
$\Lambda_{\textrm{out}}\cup P^\Lambda_{\textrm{in}}$ a strictly convex
drawing $\Lambda_{\textrm{in}}$ of~$H_{\textrm{in}}$ in which
$C_{\textrm{in}}$ is represented by~$P^\Lambda_{\textrm{in}}$, as
in~\cref{fig:operation-3-2c}; this
drawing can be constructed again by means
of~\cite{hn-cdhpgcpg-10,t-hdg-63}.  This results in a planar
straight-line drawing~$\Lambda$ of~$H$.

We now describe the morph that occurs in Operation~3. We
first define a morph $\langle \Psi,\dots,\Phi \rangle$ from $\Psi$ to
another planar straight-line drawing~$\Phi$ of~$H$, as the
concatenation of two morphs $\langle \Psi,\dots,\Lambda \rangle$ and
$\langle \Lambda,\dots,\Phi \rangle$.  The morph
$\langle \Psi,\dots,\Lambda \rangle$ is an $O(n)$-step crossing-free 2D morph obtained by applying the algorithm of Alamdari et al.~\cite{aaccddfhlprsw-hmpgd-SICOMP17}.  The morph
$\langle \Lambda,\dots,\Phi \rangle$ is an $O(1)$-step
3D morph that is obtained by applying Operation~1 to
$\Lambda_{\textrm{in}}$ only, with $u$ and $v$ fixed;
\cref{fig:operation-3-2d} shows the resulting drawing $\Phi$. In order to
prove that Operation~3 defines a crossing-free morph, it suffices to
observe that, during $\langle \Lambda,\dots,\Phi \rangle$, the
intersection of $H_{\textrm{in}}$ with the plane on which
$\Lambda_{\textrm{out}}$ lies is (a subset of) the segment
$\overline{uv}$, which lies in the interior of a face of
$\Lambda_{\textrm{out}}$; hence, $H_{\textrm{in}}$ does not cross
$H_{\textrm{out}}$.  That no other crossings occur during
$\langle \Psi,\dots,\Phi \rangle$ is a consequence of the results of
Alamdari et al.~\cite{aaccddfhlprsw-hmpgd-SICOMP17} (which ensure that
$\langle \Psi,\dots,\Lambda \rangle$ has  no crossings) and of
the properties of Operation~1 (which ensure that $\langle \Lambda,\dots,\Phi \rangle$ has no
crossings between edges of $H_{\textrm{out}}$).  Finally, Operation~3 is the
morph $\langle \Gamma,\dots,\Gamma' \rangle$ obtained by restricting
the morph $\langle \Psi,\dots,\Phi \rangle$ to the vertices and edges of~$G$. Note that the effect of Operation~1, applied only
to~$\Lambda_{\textrm{in}}$, is the one of flipping the embeddings of
$G_i,\dots,G_j$ (and also reversing their order around $u$),
while leaving the embeddings of $G_1,\dots,G_{i-1},G_{j+1},\dots,G_{k}$ unaltered, as claimed.


\smallskip

\begin{figure}[tb]
  \begin{subfigure}[b]{.24\linewidth}
    \centering
    \includegraphics[page=1,scale=0.65]{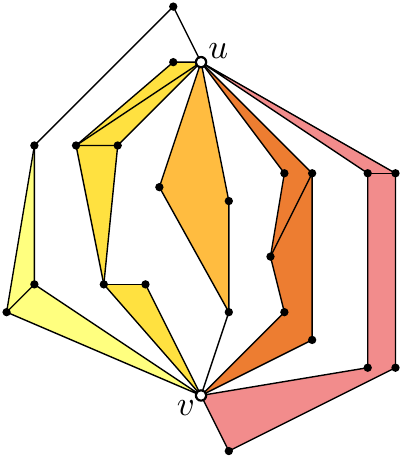}
  \end{subfigure}
  \hfill
  \begin{subfigure}[b]{.25\linewidth}
    \centering
    \includegraphics[page=2,scale=0.65]{Operation4}
  \end{subfigure}
  \hfill
  \begin{subfigure}[b]{.22\linewidth}
    \centering
    \includegraphics[page=4,scale=0.65]{Operation4}
  \end{subfigure}
  \hfill
  \begin{subfigure}[b]{.2\linewidth}
    \centering
    \includegraphics[page=3,scale=0.65]{Operation4}
  \end{subfigure}

  \vspace{-1ex}
  
  \begin{subfigure}[t]{.24\linewidth}
    \centering
    \caption{Drawing $\Gamma$ of~$G$.}
    \label{fig:operation-4a}
  \end{subfigure}
  \hfill
  \begin{subfigure}[t]{.25\linewidth}
    \centering
    \caption{Drawing $\Psi$ of $H$.}
    \label{fig:operation-4b}
  \end{subfigure}
  \hfill
  \begin{subfigure}[t]{.22\linewidth}
    \centering
    \caption{Drawing
        $\Lambda_{\textrm{out}}\cup P^\Lambda_{\textrm{in}}$ of
        $H_{\textrm{out}}\cup C_{\textrm{in}}$.}
    \label{fig:operation-4c}
  \end{subfigure}
  \hfill
  \begin{subfigure}[t]{.2\linewidth}
    \centering
    \caption{Drawing $\Lambda$ of~$H$.}
    \label{fig:operation-4d}
  \end{subfigure}

  \vspace{-1ex}
              
  \caption{Illustration for Operation~4: $P_{\textrm{in}}$ is blue,
    $P_{\textrm{out}}$ is red, and $P_{\textrm{ext}}$ is purple.}
  \label{fig:operation-4}
\end{figure}

\noindent
{\bf Operation 4: Component skip.} Operation~4 works in a setting
similar to the one of Operation~3. Specifically, $G$, $G_1,\dots,G_k$,
$\{u,v\}$, and $\Gamma$ are defined as in Operation 3; see
\cref{fig:operation-4a}.
However, we have one further assumption: If the edge $(u,v)$ exists,
then it is the split component $G_1$. Let $i$ be any index in
$\{2,\dots,k\}$. Operation~4 allows $G_i$ to ``skip'' the other components of $G$, so to be incident to the outer face. This is formalized in the following.  
\begin{lemma}\label{le:op4}
There exists an $O(n)$-step 3D crossing-free morph from $\Gamma$ to a
planar straight-line drawing $\Gamma'$ in which the embedding of $G_\ell$
is the same as in $\Gamma$, for $\ell=1,\dots,k$, and the clockwise order
of the split components around $u$ is
$G_1,\dots,G_{i-1},$ $G_{i+1}$, $\dots,G_k,G_i$, where $G_1$ and $G_i$ are
incident to the outer face.
\end{lemma}


In order to implement Operation~4, which proves \cref{le:op4}, we would like to first move $G_i$ vertically up from the plane $z=0$ to the plane $z=1$, to then send $G_i$ ``far away'' by modifying the $x$- and $y$-coordinates of its vertices, and to finally project $G_i$ vertically back to the plane $z=0$. There are two complications to this plan, though. The first one is given by the vertices $u$ and $v$, which belong both to $G_i$ and to the rest of the graph. When moving $u$ and $v$ on the plane $z=1$, the edges incident to them are dragged along, which might result in these edges crossings each other. The second one is that there might be no far away position that allows the drawing of $G_i$ to be vertically projected back to the plane $z=0$ without introducing any crossings. This is because the rest of the graph might be arbitrarily mingled with $G_i$ in the initial drawing $\Gamma$. As in Operation~3, convexity comes to the rescue. Indeed, we first employ a 2D crossing-free morph which makes the boundary of the outer face of $G$ convex and moves $G_i$ into a convex polygon. After moving $G_i$ vertically up to the plane $z=1$, sending $G_i$ far away can be simply implemented as a scaling operation, which ensures that the edges incident to $u$ and $v$ do not cross each other during the motion of $G_i$ on the plane $z=1$ and that projecting $G_i$ vertically back to the plane $z=0$ does not introduce crossings with the edges of the rest of the graph.

We now provide the details of Operation~4, which works slightly differently if the edge
$(u,v)$ exists and if it does not. We first describe
the latter case.
Refer to~\cref{fig:operation-4b}.
We insert two polygons $P_{\textrm{in}}$ and
$P_{\textrm{out}}$ with $O(n)$ vertices in~$\Gamma$. As in
Operation~3, they intersect $\Gamma$ only at $u$ and $v$, with
$P_{\textrm{in}}$ inside $P_{\textrm{out}}$ (except at~$u$ and~$v$).
All the vertices of $G_i$ (except~$u$
and~$v$) lie inside $P_{\textrm{in}}$ and all the vertices of
$G_1,\dots,G_{i-1},G_{i+1},\dots,G_k$ (except $u$ and $v$) lie
outside $P_{\textrm{out}}$. We also insert in $\Gamma$ a polygon
$P_{\textrm{ext}}$, with $O(n)$ vertices, that intersects $\Gamma$ only at $u$ and $v$, and that contains all the
vertices of $G$ and $P_{\textrm{out}}$ (except $u$ and $v$) in its
interior.

We now triangulate the region inside $P_{\textrm{ext}}$ and outside
$P_{\textrm{out}}$, without introducing chords for
$P_{\textrm{out}}$.  We also triangulate the
interior of~$P_{\textrm{in}}$ without introducing
chords for~$P_{\textrm{in}}$.  Let~$\Psi$ be the resulting planar straight-line drawing of a plane
graph~$H$.  Let $C_{\textrm{out}}$, $C_{\textrm{in}}$, and
$C_{\textrm{ext}}$ be the cycles of $H$ represented by
$P_{\textrm{out}}$, $P_{\textrm{in}}$, and $P_{\textrm{ext}}$ in
$\Psi$, respectively, and let $H_{\textrm{out}}$ ($H_{\textrm{in}}$)
be the subgraph of $H$ induced by the vertices that lie outside or on
$P_{\textrm{out}}$ (resp.\ inside or on $P_{\textrm{in}}$).  Note that
$H_{\textrm{out}}$ is an internally-triconnected plane graph and $H_{\textrm{in}}$ is a triconnected plane graph.

We now construct another planar straight-line drawing of $H$, as
follows. First, construct a strictly convex drawing~$Q_{\textrm{ext}}$
of~$C_{\textrm{ext}}$ such that the angle of~$Q_{\textrm{ext}}$ at~$u$
(and the angle at $v$) is cut by the segment $\overline{uv}$ into two
angles both smaller than $90^\circ$.  Next, construct a
strictly convex drawing~$\Lambda_{\textrm{out}}$ of~$H_{\textrm{out}}$
in which $C_{\textrm{ext}}$ is represented by $Q_{\textrm{ext}}$, by
means of~\cite{hn-cdhpgcpg-10,t-hdg-63}. Let~$P^\Lambda_{\textrm{out}}$ be the strictly convex polygon
representing~$C_{\textrm{out}}$ in~$\Lambda_{\textrm{out}}$.
As in~\cref{fig:operation-4c}, plug a strictly convex drawing
$P^\Lambda_{\textrm{in}}$ of $C_{\textrm{in}}$ in the interior of $P^\Lambda_{\textrm{out}}$, except at $u$ and $v$, so that the path $\mathcal{P}_{\textrm{in}}$ that is traversed when walking in clockwise direction along $C_{\textrm{in}}$ from $u$ to $v$ is represented by the straight-line segment $\overline{uv}$. 
Finally, plug into $\Lambda_{\textrm{out}}\cup P^\Lambda_{\textrm{in}}$
a convex drawing~$\Lambda_{\textrm{in}}$ of~$H_{\textrm{in}}$ in which
the outer face is delimited by~$P^\Lambda_{\textrm{in}}$, by means of~\cite{hn-cdhpgcpg-10,t-hdg-63}.
This results in a planar straight-line drawing $\Lambda$ of $H$,
see \cref{fig:operation-4d}.


We now describe the morph that occurs in Operation~4.  We
first define a morph $\langle \Psi,\dots,\Phi \rangle$ from $\Psi$ to
an ``almost'' planar straight-line drawing $\Phi$ of~$H$, as the
concatenation of two morphs $\langle \Psi,\dots,\Lambda \rangle$ and
$\langle \Lambda,\dots,\Phi \rangle$.  The morph
$\langle \Psi,\dots,\Lambda \rangle$ is an $O(n)$-step crossing-free 2D morph obtained by applying the algorithm in~\cite{aaccddfhlprsw-hmpgd-SICOMP17}.  Translate and rotate the
Cartesian axes so that, in $\Lambda$, the $y$-axis passes through~$u$
and~$v$ and $u$ has a smaller $y$-coordinate than $v$.  The morph
$\langle \Lambda,\dots,\Phi \rangle$ is a $3$-step 3D
morph defined as follows.
\begin{itemize}
	\item The first morphing step $\langle \Lambda,\Lambda' \rangle$ moves all the vertices of $H_{\textrm{in}}$, except for $u$ and $v$, vertically up, to the plane $z=1$. As the projection to the plane $z=0$ of every drawing of $H$ in $\langle \Lambda,\Lambda' \rangle$ coincides with $\Lambda$, the morph is crossing-free.
	\item The second morphing step $\langle \Lambda',\Lambda'' \rangle$ is such that $\Lambda''$ coincides with $\Lambda'$, except for the $x$-coordinates of the vertices of $H_{\textrm{in}}$, which are all multiplied by the same real value $s>0$. The value $s$ is large enough so that, in $\Lambda''$, the following property holds true: The absolute value of the slope of the line through $u$ and through the projection to the plane $z=0$ of any vertex of $H_{\textrm{in}}$ not in $\mathcal P_{\textrm{in}}$ is smaller than the absolute value of the slope of every edge incident to $u$ in $H_{\textrm{out}}$; and likewise with $v$ in place of $u$. This morph is crossing-free, as it just scales the drawing of $H_{\textrm{in}}$ up, while leaving the drawing of $H_{\textrm{out}}$ unaltered. Intuitively, this is the step where $G_i$ ``skips'' $G_{i+1},\dots,G_k$ (although it still lies on a different plane than those components, except for $u$ and $v$). 	
	\item The third morphing step $\langle \Lambda'',\Phi \rangle$ moves the vertices of $H_{\textrm{in}}$ vertically down, to the plane $z=0$. This morphing step might actually have crossings in its final drawing $\Phi$. However, the property on the slopes guaranteed by the second morphing step ensures that the only crossings are those involving edges incident to vertices of $\mathcal P_{\textrm{in}}$ different from $u$ and $v$, which do not belong to $G$. Hence, the restriction of $\langle \Lambda'',\Phi \rangle$ to $G$ is a crossing-free morph. 
\end{itemize}

\begin{figure}[tb]
  \begin{subfigure}[b]{.3\linewidth}
    \centering
    \includegraphics[page=5,scale=0.65]{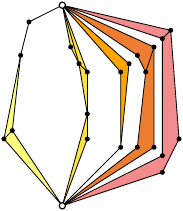}
  \end{subfigure}
  \hfill
  \begin{subfigure}[b]{.62\linewidth}
    \centering
    \includegraphics[page=6,scale=0.65]{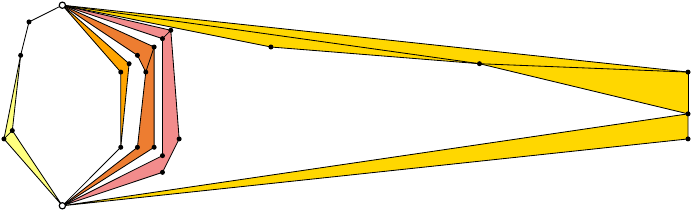}
  \end{subfigure}

  \vspace{-1ex}
  
  \begin{subfigure}[t]{.3\linewidth}
    \caption{Drawing obtained by restricting $\Lambda$ to~$G$.}
    \label{fig:operation-4-3a}
  \end{subfigure}
  \hfill
  \begin{subfigure}[t]{.62\linewidth}
    \caption{Drawing $\Gamma'$ obtained by restricting~$\Phi$ to~$G$.}
    \label{fig:operation-4-3b}
  \end{subfigure}

  \vspace{-1ex}
              
  \caption{Illustration for Operation~4: Construction of $\Gamma'$ from the restriction of $\Lambda$ to~$G$.}
  \label{fig:operation-4-3}
\end{figure}

As in Operation~3, the actual planar morph
$\langle \Gamma,\dots,\Gamma' \rangle$ is obtained by restricting the
morph $\langle \Psi,\dots,\Phi \rangle$ to~$G$, see
\cref{fig:operation-4-3}.

We now discuss the case that the edge $(u,v)$ exists; then~$G_1$ is such an edge.  Now~$P_{\textrm{in}}$
and~$P_{\textrm{out}}$ surround all the components $G_1,\dots,G_i$,
and not just~$G_i$; consequently, $H_{\textrm{in}}$ comprises $G_1,\dots,G_i$.  The description of Operation~4
remains the same, except for two differences. First, $P^{\Lambda}_{\textrm{in}}$ is strictly convex; in particular, $\mathcal P_{\textrm{in}}$ is not represented by a straight-line segment, so that the edge $(u,v)$ lies in the interior of $P^{\Lambda}_{\textrm{in}}$. Second, in the $3$-step 3D morph
$\langle \Lambda,\Lambda',\Lambda'',\Phi \rangle$, not all the
vertices of $H_{\textrm{in}}$ are lifted to the plane $z=1$, then
scaled, and then projected back to the plane $z=0$, but only those
of~$G_i$.  The arguments for the fact that the restriction of such a
morph to~$G$ is crossing-free remain the same.

\subsection{3D Morphs for Biconnected Planar Graphs} \label{sse:biconnected}

We now describe an algorithm that constructs an $O(n^2)$-step morph between any two planar straight-line drawings $\Gamma$ and $\Phi$ of the same $n$-vertex biconnected planar graph $G$. It actually suffices to construct an $O(n^2)$-step morph from $\Gamma$ to {\em any} planar straight-line drawing $\Lambda$ of $G$ with {\em the same embedding} as $\Phi$, as then an $O(n)$-step morph from $\Lambda$ to $\Phi$ can be constructed by means of~\cite{aaccddfhlprsw-hmpgd-SICOMP17}. And even more, it suffices to construct an $O(n^2)$-step morph from $\Gamma$ to {\em any} planar straight-line drawing $\Psi$ of $G$ that has {\em the same rotation system} as $\Lambda$, as then an $O(1)$-step morph from $\Psi$ to $\Lambda$ can be constructed by means of Operation 2. 

As proved by Di Battista and Tamassia~\cite{dt-opt-96}, starting from a planar graph drawing (in our case, $\Gamma$), one can obtain the rotation system of any other planar drawing (in our case, $\Phi$) of the same graph by: (i) suitably changing the permutation of the components in some \emph{parallel compositions}; that is, for some split pairs $\{u,v\}$ that define three or more split components, changing the clockwise (circular) ordering of such components; and (ii) flipping the embedding for some \emph{rigid compositions}; that is, for some split pairs that define a maximal split component that is biconnected, flipping the embedding of the component. Thus, it suffices to show how to implement these modifications by means of Operations 2--4 from Section~\ref{sse:operations}. We  first take care of the flips, not only in the description, but also algorithmically:
All the flips are performed before all the permutation rearrangements since the flips might cause some permutation changes, which we then fix later.


Let $\{u,v\}$ be a split pair that defines a maximal biconnected split component $K$ of $G$, and suppose that we want to flip the embedding of $K$ in $\Gamma$ (the drawing we deal with undergoes modifications, however for the sake of simplicity we always denote it by $\Gamma$). Note that $K$ is not the edge $(u,v)$, as otherwise we would not need to flip its embedding. Further, $(u,v)$ does not belong to $K$, as otherwise $K$ would not be a maximal split component. However, $(u,v)$ might belong to $E(G)-E(K)$. Apply Operation 2 to morph $\Gamma$ so that the outer face becomes any face incident to $u$ and $v$. Let $G_1,\dots,G_k$ be the split components of $G$ with respect to $\{u,v\}$, in clockwise order around $u$, where $G_1$ and $G_k$ are incident to the outer face. Let $\ell\in\{1,\dots,k\}$ be such that $G_\ell=K$. We distinguish two cases, depending on whether the edge $(u,v)$ belongs to $G$ or not. 

\begin{itemize}
	\item If the edge $(u,v)$ does not belong to $G$, then we simply apply Operation~3, with $i=j=\ell$, in order to morph $\Gamma$ to flip the embedding of $G_\ell=K$.
	\item If $(u,v)$ belongs to $G$, then let $\emm\in\{1,\dots,k\}$ be such that $G_\emm$ is $(u,v)$. Assume that $\ell<\emm$, the other case is symmetric. Apply Operation~3 with $i=\ell$ and $j=\emm$, in order to morph $\Gamma$ to flip the embeddings of $G_\ell,G_{\ell+1},\dots,G_\emm$. If we again denote by $G_1,\dots,G_k$ the split components of $G$ with respect to $\{u,v\}$, in clockwise order around $u$, where $G_1$ and $G_k$ are incident to the outer face, $G_\ell$ is now the edge $(u,v)$ and $G_\emm$ is $K$. Apply Operation~3 a second time, with $i=\ell$ and $j=\emm-1$, in order to morph $\Gamma$ to flip the embeddings of $G_\ell,G_{\ell+1},\dots,G_{\emm-1}$ back to the embeddings they originally had. As desired, only the embedding of $K$ is actually flipped.
\end{itemize}

Flipping the embedding of $K$ is hence done in $O(n)$ morphing steps. Since there are $O(n)$ maximal biconnected split components whose embedding might need to be flipped, all such flips are performed in $O(n^2)$ morphing steps.

Let  $\{u,v\}$ be a split pair of $G$ that defines three or more split components and suppose that we want to change the clockwise (circular) ordering of such components around $u$ to a different one. If the edge $(u,v)$ exists, then apply Operation 2 to morph $\Gamma$ so that the outer face becomes the one to the left of $(u,v)$, when traversing $(u,v)$ from $u$ to $v$; otherwise, apply Operation 2 to morph $\Gamma$ so that the outer face becomes any face incident to $u$ and $v$. Let $G_1,\dots,G_k$ be the split components of $G$ with respect to $\{u,v\}$, in clockwise order around $u$, where $G_1$ and $G_k$ are incident to the outer face; note that, if $(u,v)$ exists, then it coincides with $G_1$. Let $G_1,G_{\sigma(2)},G_{\sigma(3)},\dots,G_{\sigma(k)}$ be the desired clockwise order of the split components of $G$ with respect to $\{u,v\}$ around $u$; since we are only required to fix a clockwise {\em circular} order of these components, then we can assume $G_1$ to be the first component in the desired clockwise {\em linear} order of such components around $u$ that starts at the outer face.  

We apply Operation 4 with index $\sigma(2)$, then again with index $\sigma(3)$, and so on until the index $\sigma(k)$. The first $j$ applications make $G_{\sigma(2)},G_{\sigma(3)},\dots,G_{\sigma(j+1)}$ the last $j$ split components of $G$ with respect to $\{u,v\}$ in the clockwise linear order of the components around $u$ that starts at the outer face. Hence, after the last application we obtain the desired order. Each application of Operation 4 requires $O(n)$ morphing steps, hence changing the clockwise order around $u$ of the split components of $G$ with respect to a split pair $\{u,v\}$ takes $O(nk)$ morphing steps, where $k$ is the number of split components with respect to $\{u,v\}$. Since the total number of split components with respect to every split pair of $G$ that defines a parallel composition is in $O(n)$~\cite{dt-opt-96}, this sums up to $O(n^2)$ morphing steps. This concludes the proof of Theorem~\ref{thm:ub-3d} for biconnected planar graphs.


\subsection{3D Morphs for General Planar Graphs} \label{sse:general}

We start by reducing the general problem to the one in which $G$ is connected. Suppose that $G$ has multiple connected components $G_1,\dots,G_k$. Assume that, for $i=1,\dots,k$, we know how to construct a 3D crossing-free morph $\mathcal M_i=\langle \Phi_{i,1},\dots,\Phi_{i,2}\rangle$ between any two planar straight-line drawings $\Phi_{i,1}$ and $\Phi_{i,2}$ of a connected component $G_i$ of $G$. Suppose that, if $\Phi_{i,1}$ and $\Phi_{i,2}$ share a point $p$, then $\mathcal M_i$ has \emph{extension} $\mathcal W_i$; that is, the entire morph $\mathcal M_i$ happens within a ball centered at $p$ with radius $\mathcal W_i$. Clearly, this is true for a sufficiently large value $\mathcal W_i>0$. Let $\mathcal W=\max_{i=1,\dots,k} \mathcal W_i$. 
Let $\Gamma_1$ and $\Gamma_2$ be the two given planar straight-line drawings of $G$ between which we want to construct a 3D crossing-free morph.  A constant number of morphing steps can be used in order to move the connected components of $G$ ``sufficiently far apart'' from one another. This is done as follows. For $j=1,2$, let $\langle \Gamma_j,\Psi_j\rangle$ be the 3D crossing-free morph that moves the drawing of $G_i$ vertically up, to the plane $z=i$; now distinct connected components of $G$ lie on different horizontal planes. For $j=1,2$, let $\langle \Psi_j,\Lambda_j\rangle$ be the 3D crossing-free morph that translates the drawing of $G_i$ on the plane $z=i$ so that it contains the point $(3i\mathcal W,0,i)$; now distinct connected components of $G$ are ``far apart''. Finally, for $j=1,2$, let $\langle \Lambda_j,\Phi_j\rangle$ be the 3D crossing-free morph that moves the drawing of $G_i$ vertically down, to the plane $z=0$. For $i=1,\dots,k$, let $\Phi_{i,j}$ be the restriction of $\Phi_j$ to $G_i$. Note that $\Phi_{i,1}$ and $\Phi_{i,2}$ share the point $(3i\mathcal W,0,0)$. This and the distance between distinct connected components of $G$ in $\Phi_1$ and $\Phi_2$ ensure that the union of crossing-free morphs $\langle \Phi_{i,1},\dots,\Phi_{i,2}\rangle$ gives us a crossing-free morph $\langle \Phi_1,\dots,\Phi_2\rangle$, and thus $\langle \Gamma_1,\Psi_1,\Lambda_1,\Phi_1,\dots,\Phi_2,\Lambda_2,\Psi_2,\Gamma_2\rangle$ is the desired morph between $\Gamma_1$ and $\Gamma_2$.

We now assume that $G$ is connected; let $\Gamma$ and $\Phi$ be the prescribed planar straight-line drawings of $G$ we want to morph. We are going to augment $\Gamma$ and $\Phi$ to planar straight-line drawings of a biconnected planar graph and then apply the algorithm of Section~\ref{sse:biconnected}. The augmentation is done in $k-1$ steps, where $k$ is the number of biconnected components of $G$. At each step, the augmentation decreases by one the number of biconnected components of $G$ by employing $O(n)$ morphing steps. Thus, the total number of morphing steps used by the augmentation is in $O(n^2)$. We now describe how a single augmentation step is done (the drawing $\Gamma$ and the graph $G$ we deal with undergo some modifications, however for the sake of simplicity we always denote them by $\Gamma$ and $G$).

\begin{figure}[tb]
	\begin{subfigure}[b]{.21\linewidth}
		\centering
		\includegraphics[page=1,scale=0.75]{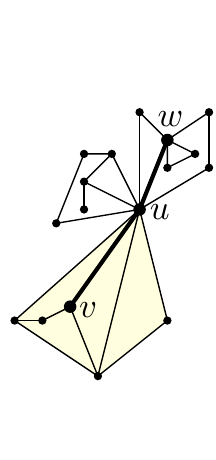}
	\end{subfigure}
	\hfill
	\begin{subfigure}[b]{.23\linewidth}
		\centering
		\includegraphics[page=2,scale=0.75]{Connected}
	\end{subfigure}
	\hfill
	\begin{subfigure}[b]{.24\linewidth}
		\centering
		\includegraphics[page=3,scale=0.75]{Connected}
	\end{subfigure}
	\hfill
	\begin{subfigure}[b]{.22\linewidth}
		\centering
		\includegraphics[page=4,scale=0.75]{Connected}
	\end{subfigure}
	
	\vspace{-1ex}
	
	\begin{subfigure}[t]{.21\linewidth}
		\caption{Drawing $\Gamma$; block~$B$ is yellow.}
		\label{fig:connected-a}
	\end{subfigure}
	\hfill
	\begin{subfigure}[t]{.23\linewidth}
		\caption{Triangulating $\Gamma$ into a drawing $\Psi$.}
		\label{fig:connected-b}
	\end{subfigure}
	\hfill
	\begin{subfigure}[t]{.24\linewidth}
		\centering
		\parbox[t]{.87\linewidth}{%
			\caption{Changing the outer face of $\Psi$.}
			\label{fig:connected-c}}
	\end{subfigure}
	\hfill
	\begin{subfigure}[t]{.22\linewidth}
		\caption{Drawing $\Gamma'$.}
		\label{fig:connected-d}
	\end{subfigure}
	
	\vspace{-1ex}
	
	\caption{Illustration for the morph that allows the path $(v,p,w)$ to be inserted in  $\Gamma$.}
	\label{fig:connected}
\end{figure}

Let $B$ be a biconnected component of $G$ that contains a unique cut-vertex $u$ (that is, $B$ is a leaf of the block-cut-vertex tree of $G$~\cite{h-gt-69,ht-aeagm-73}). Let $(u,v)$ and $(u,w)$ be two edges that are consecutive in the clockwise order of the edges incident to $u$ in $\Phi$ and such that $(u,v)\in E(B)$ and $(u,w)\notin E(B)$. We are going to augment $G$ with a length-$2$ path $(v,p,w)$, thus decreasing the number of biconnected components of $G$. Such a path can be planarly inserted in $\Phi$, because of the way $v$ and $w$ were defined. However, $v$ and $w$ are not necessarily incident to the same face of $\Gamma$, as in \cref{fig:connected-a}; in order to allow for a planar insertion of the path $(v,p,w)$, we are going to let $v$ and $w$ share a face by suitably morphing $\Gamma$.

Triangulate $\Gamma$ into a planar straight-line drawing $\Psi$ of a
maximal planar graph $H$, as in \cref{fig:connected-b}, and then apply
Operation~2 to morph $\Psi$ in $O(1)$ steps to change its outer face
into any of the two faces incident to the edge $(u,w)$, as in
\cref{fig:connected-c}; let $q$ be the third vertex incident to such a
face. By means of~\cite{hn-cdhpgcpg-10,t-hdg-63}, we construct a
planar straight-line drawing $\Lambda$ of $H$ in which the cycle
$(u,w,q)$ delimiting the outer face is represented by a triangle whose
angle at $u$ is smaller than $45^\circ$. An $O(n)$-step
crossing-free 2D morph from $\Psi$ to $\Lambda$ can be obtained
by the algorithm
in~\cite{aaccddfhlprsw-hmpgd-SICOMP17}. Restricting such morphs to $G$
provides an $O(n)$-step crossing-free morph from $\Gamma$ to a planar
straight-line drawing~$\Gamma'$ of~$G$ contained inside a triangle
$(u,w,q)$ whose angle at $u$ is smaller than $45^\circ$, as in
\cref{fig:connected-d}.

Translate and rotate the Cartesian axes so that the origin is at $u$ and the positive $y$-half-axis cuts the interior of the face that is to the right of the edge $(u,v)$, when traversing such an edge from $u$ to $v$. We are now ready to make $u$ and $v$ incident to the same face in $\Gamma'$. This is done in three morphing steps.
\begin{itemize}
	\item The first morphing step moves all the vertices of $B$, except for $u$, vertically up, to the plane $z=1$.
	\item The second morphing step scales the $x$ and $y$-coordinates of all the vertices of $B$ by a vector $(\alpha,\beta)$, where $\alpha$ and $\beta$ are two positive real values satisfying the following properties: (i) $\beta$ is sufficiently large so that every vertex in $V(B)-\{u\}$ has a $y$-coordinate larger than the one of every vertex in $V(G)-V(B)$; and (ii) $\alpha$ is large enough so that the slope of every edge $(u,r)$ of $B$ is either between $0^\circ$ and $45^\circ$ (if $r$ has positive $x$-coordinates) or between $135^\circ$ and $180^\circ$ (if $r$ has negative $x$-coordinates).
	\item The third morphing step moves all the vertices of $B$ vertically down, back to the plane $z=0$.
\end{itemize}

\begin{figure}[htb]
	\centering
	\includegraphics[page=5,scale=0.65]{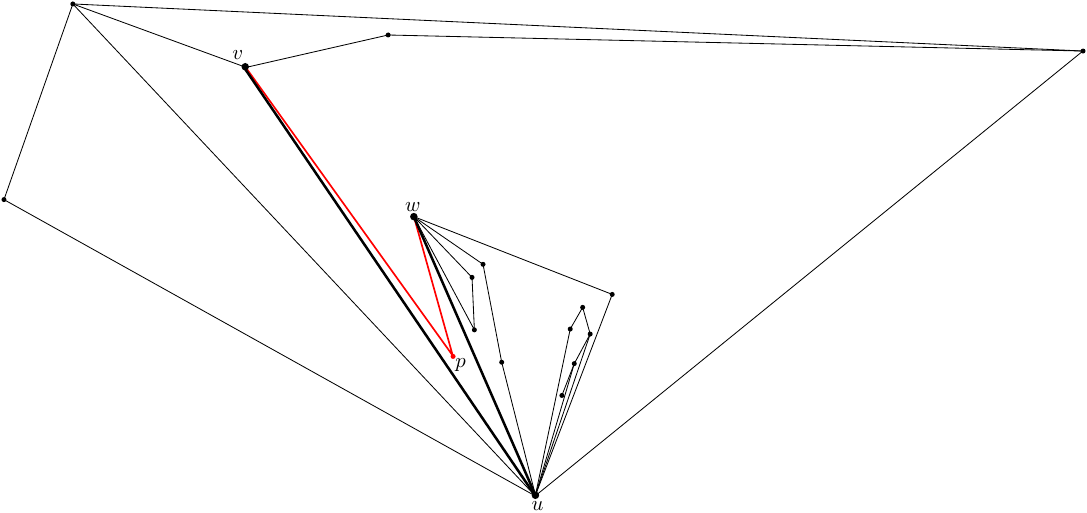}
	\caption{Illustration for the morph that allows the path $(v,p,w)$ to be inserted in  $\Gamma$. Scaling $B$ up so that it surrounds the rest of the graph.}
	\label{fig:connected-2}
\end{figure}
The first two morphing steps are clearly crossing-free. The third one is also crossing-free, because of the properties that are ensured by the choice of $\alpha$ and $\beta$ in the second morphing step. Now $v$ and $w$ are incident to the same face not only in $\Phi$, but also in $\Gamma'$. Thus, they can be connected via a length-$2$ path $(v,p,w)$; the new vertex $p$ can be inserted close to $u$, both in $\Gamma'$ and in $\Phi$, as in \cref{fig:connected-2}. Now $B$ and the biconnected component $w$ used to belong to have been merged into a single biconnected component, as desired.

\section{Discussion: Lower Bounds}
\label{sec:lower}

Though the algorithm of~\cref{sec:upper} uses a quadratic number of steps, we are not aware of any super-constant lower bound for crossing-free 3D morphs between planar straight-line graph drawings.
%
The nested-triangles graph provides a linear lower bound on the number of steps required for a crossing-free {\em 2D} morph, as proved by Alamdari et al.~\cite{aaccddfhlprsw-hmpgd-SICOMP17}.
\begin{figure}[tb]
  \centering
  \includegraphics{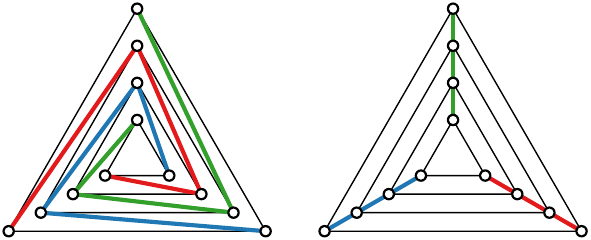}
  \caption{The lower bound example of~\cite{aaccddfhlprsw-hmpgd-SICOMP17}.}
  \label{fig:lower-bound}
\end{figure}
Specifically, let~$T_k$ be the pair of drawings of the graph that
consists of $k+1$ nested triangles, connected by three paths that are
spiraling in the first drawing and straight in the second drawing, as
in \cref{fig:lower-bound} for $k=3$.
The lower bound of~\cite{aaccddfhlprsw-hmpgd-SICOMP17} relies on the fact that the innermost triangle or the outermost triangle makes a linear number of full turns in any crossing-free 2D morph between the two drawings.

\renewcommand{\topfraction}{0.99} \renewcommand{\bottomfraction}{0.98}  \renewcommand{\textfraction}{0.01}

\begin{figure}
	\includegraphics[scale=.4,page=1]{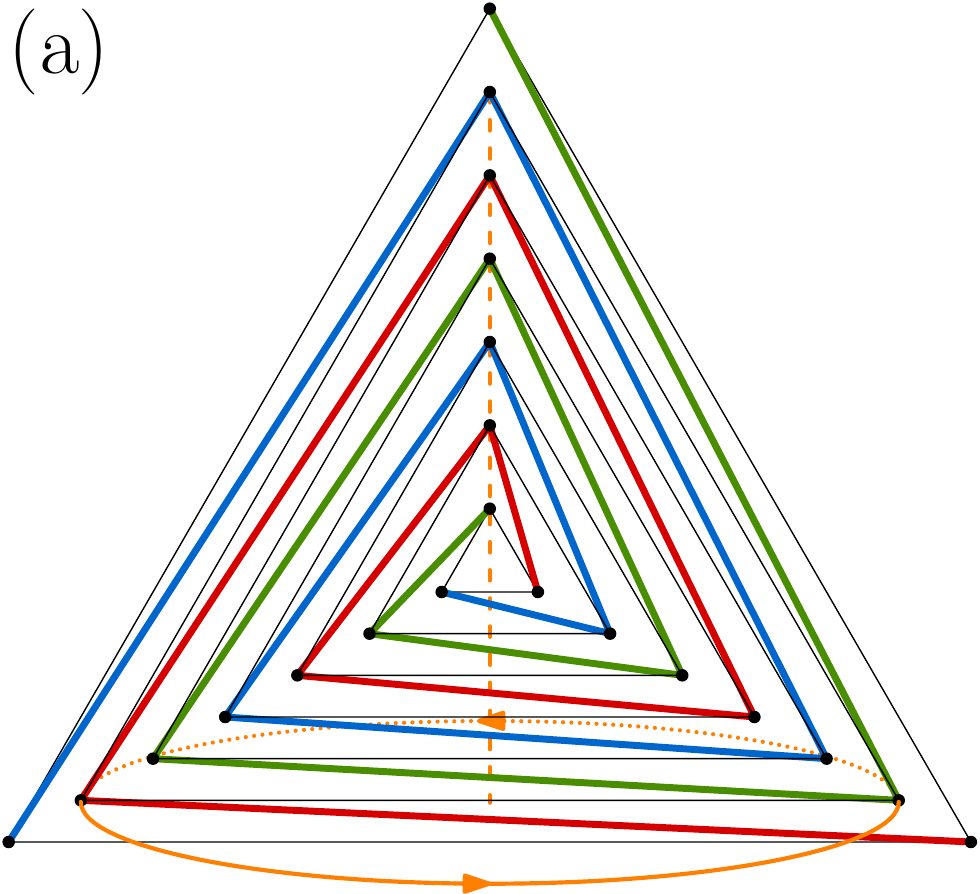}
	\includegraphics[scale=.4,page=2]{720flips_rotations}
	\includegraphics[scale=.4,page=3]{720flips_rotations}
	
	\includegraphics[scale=.4,page=4]{720flips_rotations}
	\includegraphics[scale=.4,page=5]{720flips_rotations}
	\includegraphics[scale=.4,page=6]{720flips_rotations}
	\caption{Morphing $T_6$ in~3D without moving the innermost and outermost triangles. Orange arrows show the vertices that exchange position in the next step. Empty$\,$/$\,$large disks indicate that a vertex lies below$\,$/$\,$above the plane containing the initial drawing. The drawing obtained by the morph is of the type of the right drawing in \cref{fig:lower-bound}.}
	\label{fig:spiral-3d}
\end{figure}

Even in 3D, it might seem that a linear number of linear morphs is required.
However, the extra dimension allows us to perform the ``turns'' in parallel by ``flipping'' several triangles at once.
The key operation is to morph $T_6$ in a constant number of steps without moving the innermost and outermost triangles, as shown in~\cref{fig:spiral-3d} and animated in~\cite{nt-geo,nt-you}.
Then for any $k$, we can construct a crossing-free 3D morph between the two drawings in $T_{6k}$ in a constant number of steps by performing the morph of~\cref{fig:spiral-3d} in parallel for the $k$ nested copies of $T_6$.
Observe that in this morph the $(6i+1)$-th outermost triangle does not move, for any $i=0,\dots,k$.
Each morphing step of $T_6$ avoids a small tetrahedron above and below its innermost triangle, allowing different nested copies of $T_6$ to morph in parallel without intersecting.



  The above example gives hope that the number of steps required to construct a crossing-free 3D morph between any two given planar straight-line graph drawings could be far smaller than quadratic~-- potentially even constant. However, it is unclear how to generalize our procedure.

\begin{figure}[tb]
  	\centering 
  	\includegraphics[width=.45\textwidth,page=1]{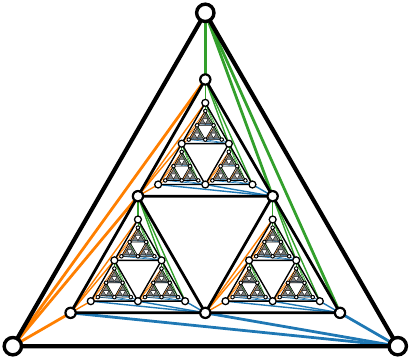}
  	\includegraphics[width=.45\textwidth,page=2]{lb_tree}
  	\caption{A potential lower bound construction.}
  	\label{fig:lb-tree}
  \end{figure}

  The approach of \cref{fig:spiral-3d} relies on the sequence of nested triangles to be {\em independent}, as we can untangle each one locally without affecting the others.
  This is not necessarily the case.  
  For instance, the example in \cref{fig:lb-tree}
  shows a tree of nested triangles that are recursively twisted by
  $120^\circ$ at every level.  Here, each path in the tree has the same
  structure as a nested-triangles graph thus, in total, it requires
  $\Omega(\log n)$ morphing steps in~2D. It is unclear to us how to
  handle the dependencies between different tree branches.

\section{Open Problems}
\label{sec:open}

Our research raises several other open problems. An immediate one is
to reduce our quadratic upper bound for the number of steps that are
needed to construct a crossing-free 3D morph between any two planar
straight-line graph drawings. Extending the result of Arseneva at
al.~\cite{abcddflt-pd3dm-JGAA19}, we ask whether planar graph families
richer than trees, e.g., outerplanar graphs and series-parallel
graphs, admit crossing-free 3D morphs with a sub-linear number of
steps.

We have given an example of two topologically equivalent planar
straight-line drawings of a triconnected graph that can be
untangled in~3D using only $O(1)$ steps.  Still we think that there
are examples of planar graphs with topologically equivalent drawings
where this is not the case.  More specifically, we suspect that in
3D, as in 2D, a linear number of steps is sometimes necessary.

\subsubsection{Acknowledgements.} The research for this paper started at the Dagstuhl Seminar 22062: ``Computation and Reconfiguration in Low-Dimensional Topological Spaces''. The authors thank the organizers and the other participants for a stimulating atmosphere and interesting discussions.

\bibliographystyle{abbrvurl}
\bibliography{morphing}

\end{document}